# Automated Smart Wick System-Based Microfarm Using Internet of Things

R. Jorda, Jr., C. Alcabasa, A. Buhay, E. C. Dela Cruz, J. P. Mendoza, A. Tolentino, L. K. Tolentino, E. Fernandez,
A. Thio-ac, J. Velasco, and N. Arago
*Technological University of the Philippines, Manila, Philippines*
romeo_jorda@tup.edu.ph

*Abstract*—This paper presents a study conducted to allow urban farmers to remotely monitor their farm through the design and development of an Internet of Things-based (IoT) microfarm prototype which utilized wick system as planting method. The system involves the detection of three environmental parameters namely, light intensity, soil moisture and temperature through the use of respective sensors which were connected to the Arduino microcontroller, the sensor node of the system. Irregularities in the aforementioned parameters were neutralized through the use of parameter regulators such as LED growlight strips, water pump and air cooler. The data collected by these sensors were gathered by the Arduino microcontroller and were sent to the Web database through the IoT gateway which was the Raspberry Pi computer chip. These data were also sent to an Android unit installed with the Microfarm Companion application which was capable of monitoring and controlling the environmental parameters observed in the microfarm. The application allows the user to view the current value of the parameter involved and to choose whether to control the parameter regulators automatically or manually. The microfarm system runs autonomously which reduces the labor required to produce healthy plants and crops. Mustard greens samples were used in testing the system. After a month of monitoring the height of the samples, it was observed that the average height of the samples is about 0.23 cm taller than the standard height. The proponents has also tested the system functionality by evaluating the sensor data log that provides the values gathered by the sensors and the turn-on times of the parameter regulators. From these data, it can be observed that whenever the values obtained by the sensors fall outside the threshold range, the parameter regulators turns on, indicating that the system is working properly.

## I. Introduction

A microfarm is a type of small-scale farm that is situated in urban or suburban areas, and is usually less than 5 acres of land [1-3]. It can be just an aquarium turned into a seed bed, a small greenhouse, or a large garden where crops of different varieties are being cultivated. With the advent of modern technology, advances in the agricultural industry were made and one of the products of these advances is the introduction of smart farming [4]. Smart farming refers to a technology that allows farmers to monitor and control the farming system which uses sensing, mobile, and big data technology [5-7].

One of the prevailing technologies that are currently being used in the implementation of smart farms is the IoT [7] technology. IoT or Internet of Things is a system of physical objects embedded with electronic devices, software, and Internet connectivity that allow them to gather, transmit, and receive data [8]. It is an environment in which living and non-living things are given with exclusive identifiers and the capability to transmit data over the Internet without human interaction. A thing, in the IoT context, can refer to an individual "with a heart monitor implant, a farm animal with a biochip transponder, an automobile that has built-in sensors or any other natural or man-made object that can be assigned an IP address and provided with the ability to transfer data over a certain network" [9].

Smart farming systems also provide farmers with valuable, real-time information about the state of the soil, harvests, equipment and other environmental factors in their fields. Many studies [10-12] were made regarding this field and more are being conducted for the purpose of developing existing systems further. In a generation of people who are too busy to grab a shovel and bury a seed in the ground, developing a system of farming that employs remote-sensing is quite vital. The researchers propose to develop a system that uses IoT technology to monitor and control several environmental parameters of a micro-farm.

## II. Methodology

### A. System Components

The system is comprised mainly of parameter detection, parameter correction, relay module, Arduino module, Raspberry pi computer, Android unit and the microfarm setting itself, as shown in Figure 1.

1) *Parameter Detection*: The system was comprised of three parameter sensors namely, light intensity, soil moisture, and temperature. These sensors were responsible for obtaining data from the surrounding environment of the microfarm which will then be uploaded to the Arduino module. For soil moisture, the sensor used was an Analog Digital Soil Moisture Sensor which passes current through the soil to measure resistance that corresponds to the moisture level. For temperature detection, the sensor used was the Temperature and Humidity DHT11 Sensor which uses an NTC temperature measurement component. Lastly, for light intensity, the sensor used was the BH1750FVI Light Intensity Sensor Module which is capable of measuring data ranging from 1-65535 lux.

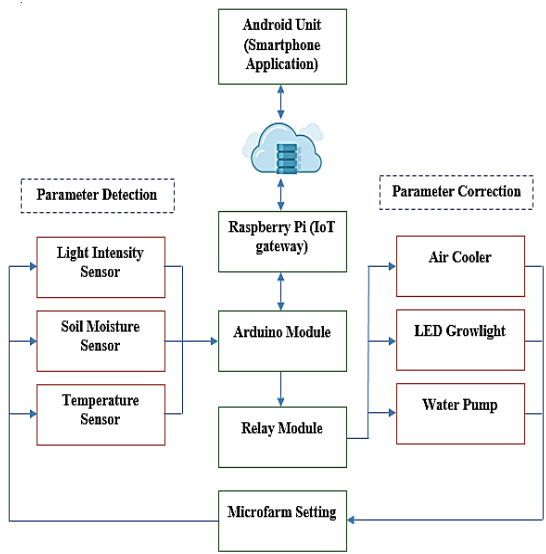

Figure 1: System Block Diagram

2) *Parameter Correction:* Three regulators, one for each sensor, were deployed in the system to correct the irregularities in parameter values. A water pump was used for maintaining the soil moisture, air cooler for the temperature and LED growlight strips for the light intensity. Depending on the user's preference, these devices can be controlled manually or automatically through the use of the Android application installed in the Android device.

3) *Relay Module:* The relay module is the component of the system that executes the command from the Android application whether to turn the parameter regulators on or off. The relay module used in the system was a single channel Relay Module SPDT (Single Pole Double Throw) that is used to control or switch devices which uses a higher voltage than what most microcontrollers such as an Arduino or Raspberry Pi can handle, making it reliable to use for controlling typical household appliances.

4) *Arduino Module:* The Arduino microcontroller serves as the sensor node of the system. The sensor node is responsible for gathering the data detected by the sensors and then forwarding it to the webserver database through the IoT gateway. It is also responsible for relaying the command sent by the smartphone application to the relay module which controls the parameter regulators. The circuit for the microfarm sensing and control is shown in Figure 2.

5) *Raspberry Pi Computer:* The Raspberry Pi acts as the IoT gateway of the system. It functions as an edge device, which obscures the sensor node from the public internet. Though the sensor node can make outbound connections to the internet and cloud through the gateway, they cannot be accessed directly. Thus, gateway plays the dual role of router and firewall securing the sensor nodes and the internal network.

6) *Android Unit:* An Android application installed in an Android device serves as a means for the user to remotely access the microfarm. The app shows the current value of the environmental parameter being measured, a Parameter Value vs. Time graph of the measured values for the last 4 hours, the date and time of the user's last visit, the value measured by that time and its percentage difference from the value being currently measured. It also provides the user with an option for automatic or manual control of the system. To avoid uncertainty, the app was programmed in such a way that when the automation feature was activated, the override feature will be disabled. Figure 3 shows the Graphical User Interface (GUI) of the app.

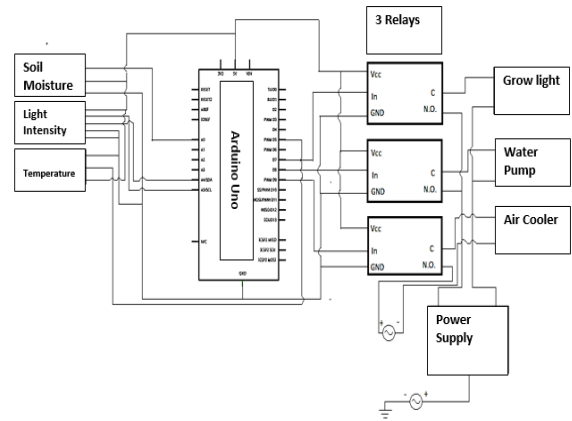

Figure 2: Pin Configuration of Arduino Module

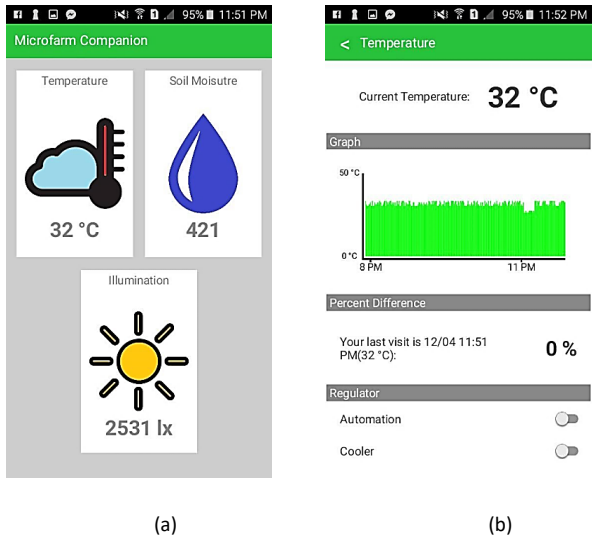

Figure 3: Android Application GUI, (a) home screen, (b) temperature screen

7) *Microfarm Setting:* This comprises of the micro-farm itself where the designated crops are being cultivated. The environmental parameters being observed herein, namely temperature, soil moisture and light intensity, will be the primary input data of the system. The primary materials used for construction of the microfarm chamber were continuous aluminum and steel brackets for the framework, and stucco-textured aluminum sheet and clear acrylic sheets for the walls. The micro-farm chamber was then mounted on a steel framework equipped with a set of castor wheels for easy transportation. The actual microfarm prototype is shown in the following figure.

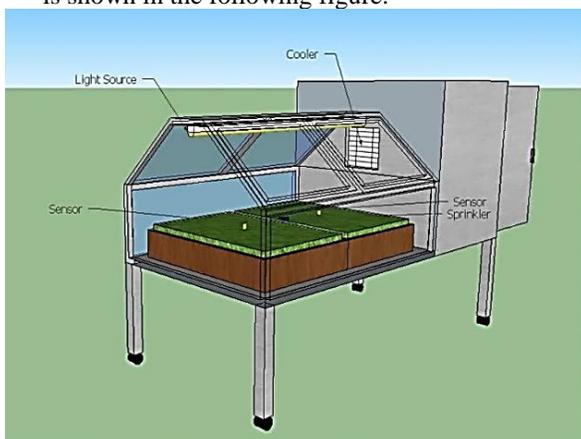

Figure 4: Design of the proposed microfarm

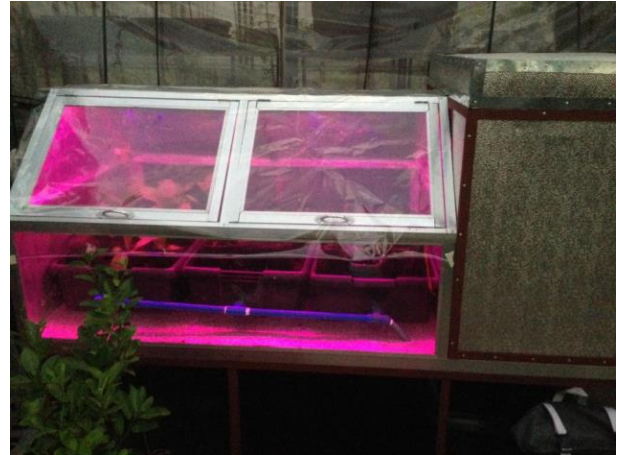

Figure 5: Actual microfarm prototype

### B. Project Technical Description

The primary input data of the system were the environmental parameters being observed inside the microfarm setting, namely temperature, soil moisture and light intensity. This data is sent to the Arduino microcontroller which is used in the system as a sensor node that gathers the data observed by the sensors and where the relay drivers that control the parameter regulators are connected. All these sensors and regulators are connected to the Raspberry Pi that serves as the gateway for reaching the cloud database. The "things" in this Internet-of-Things system are the parameter regulators since they are the ones assigned with IP addresses.

To maintain a suitable environmental condition for the plants inside the microfarm, data gathered by the sensors is sent by the Arduino every 5 seconds to the Web database. The system then processes this data to determine whether the values received is within the preset threshold values. If the values fall within the range, the system turns off the parameter regulator involved or keeps them turned off, if the regulators were turned off. If the value received falls above or below the range, the system turns on the parameter regulator involved until the desired parameter values were achieved. This is the automatic control feature of the system. However, the system also has an override feature that allows users to manually control the parameter regulators. Switching between manual and automatic control can be done using the smartphone application.

### C. Software Development

Sensor node configuration was done by programming the Arduino module using Arduino IDE which utilizes C language. The development of the Android application was done using MIT AppInventor2, and IoT Gateway to Webserver interfacing was implemented using Python IDE.

## D. Planting Method

The planting method used in the microfarm is the wick system. In this planting method, the water or the nutrient solution is injected "into the grow bed from the water tank through the capillary action of wick material and absorbent grow media" [13, 14]. In wick systems, it is highly important to consider the absorption capability of the wick to be used. Several materials such as wool, cotton, wood, soil, and gravel can be used as wicks. An advantage of this system is the passive, low-technology, low cost approach to supplying an efficient, constant source of moisture to plants as they need it. In addition, a variety of recycled materials can be used to construct the system. For this study, black decoration box was used as reservoir. A divider was used to separate the grow bed from the tank. The wick used for the system is felt cloth and the plant containers used are also made of the same material. The following figure shows a simple demonstration of the wick system.

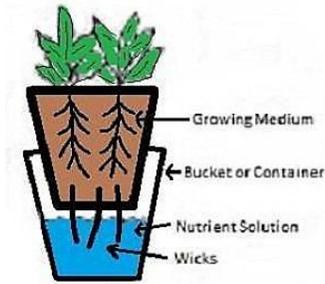

Figure 6: Wick System Using Pots

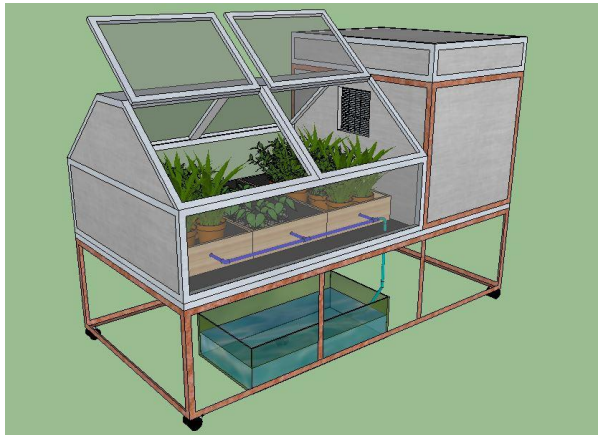

Figure 7: Wick System-Based Microfarm

## III. RESULTS AND DISCUSSION

This section shows the data gathered throughout the conduct of the study. This includes plant height monitoring and the corresponding t-test and the sensor data log.

## A. Plant Growth Monitoring

Table 1 shows the measured height of the mustard greens samples recorded by the proponents in a span of almost one month. A one-sample t-test was conducted to identify whether the height of the mustard greens samples grown inside the microfarm conforms to its standard height which is 24.688cm.

Table 1
Height Record of Mustard Greens in centimeter (cm)

| Sample Number | Day 1 9-Jan | Day 5 13-Jan | Day 15 23-Jan | Day 18 26-Jan | Day 24 1-Feb | Day 29 6-Feb |
|---|---|---|---|---|---|---|
| 1 | 8.5 | 10.1 | 16.8 | 18.7 | 22.6 | 25.1 |
| 2 | 9.8 | 13.2 | 20.5 | 23.2 | 25.2 | 26.7 |
| 3 | 9.6 | 13.5 | 19.7 | 22.1 | 23.1 | 24.9 |
| 4 | 7 | 9.9 | 15.5 | 18.3 | 21.1 | 24.4 |
| 5 | 6.8 | 7.9 | 15.1 | 17.8 | 20.4 | 23.9 |
| 6 | 8.1 | 10.5 | 15.9 | 17.4 | 20.1 | 24.5 |
| 7 | 10.5 | 13.9 | 20.3 | 22.5 | 24 | 25.1 |
| 8 | 11 | 14.2 | 21.1 | 23.5 | 25.1 | 26.3 |
| 9 | 5.4 | 7.1 | 14.8 | 16.9 | 19.3 | 22.8 |
| 10 | 8.1 | 10.6 | 16.5 | 18.4 | 22.5 | 24.8 |
| 11 | 9.5 | 12.1 | 19.6 | 22.1 | 24.3 | 25.6 |

Tables 2 and 3 shows the results of the statistical test performed on the heights of the mustard green samples. From the data, it can be observed that difference between the true mean height and the comparison value height is not equal to zero. Thus, the null hypothesis that true mean height equals comparison value height is rejected. Therefore, it can be concluded that the mean height of the sample is significantly different than the standard height of the mustard during its 27th day of growth. The average height of the sample is about 0.23 cm taller than the standard height.

Table 2
One Sample t-test Statistics for Mustard Greens

| | N | Mean | Std. Deviation | Std. Error Mean |
|---|---|---|---|---|
| Height of Mustard | 11 | 24.9182 | 1.07686 | 0.32469 |

Table 3
One Sample t-test Results for Mustard Greens

| | Test Value = 24.688 | | | | | |
|---|---|---|---|---|---|---|
| | T | df | Sig. (two-tailed) | Mean Difference | 95% Confidence Interval of the Difference | |
| | | | | | Lower | Upper |
| Height of Mustard | 0.709 | 10 | 0.495 | 0.23018 | -0.4933 | 0.9536 |

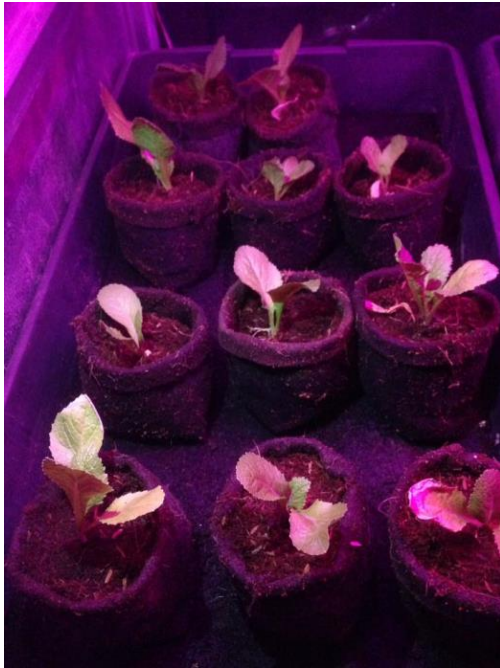

Figure 7: Mustard Greens Grow Bed Using Felt Cloth as Wick

### B. *Sensor Data Log*

The following figures show graphs representing the values gathered by the sensors as well as the turn-on times of the parameter regulators on January 9, 2017. From the graphs, it can be observed that during the time that the parameter values fall above or below the threshold value, the regulators turn on.

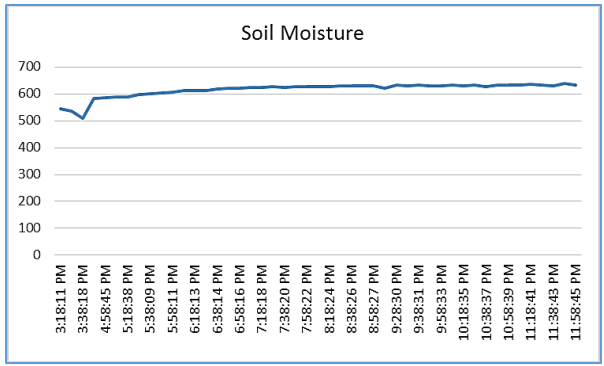

Figure 7: Graph of Sensor Data Log and Turn-On Time of Soil Moisture Regulators on Day 1

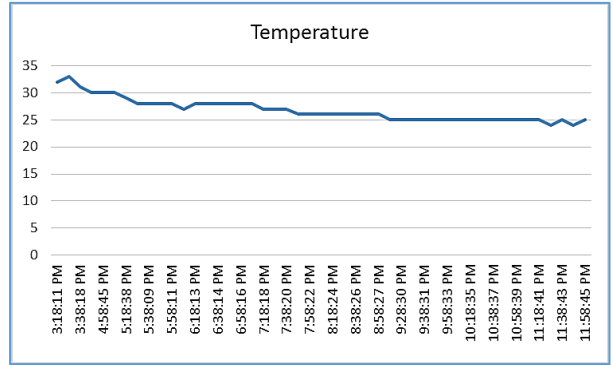

Figure 8: Graph of Sensor Data Log and Turn-On Time of Temperature Regulators on Day 1

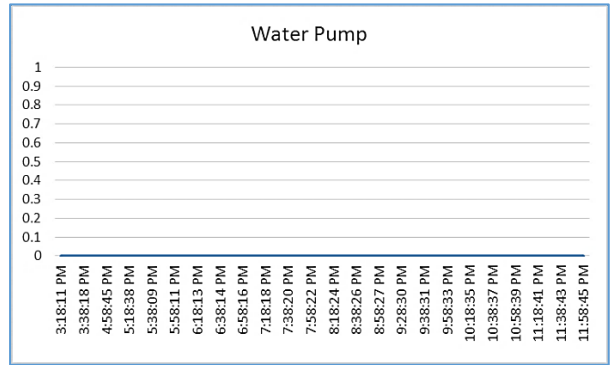

Figure 9: Graph of Sensor Data Log and Turn-On Time of Water Pump Regulators on Day 1

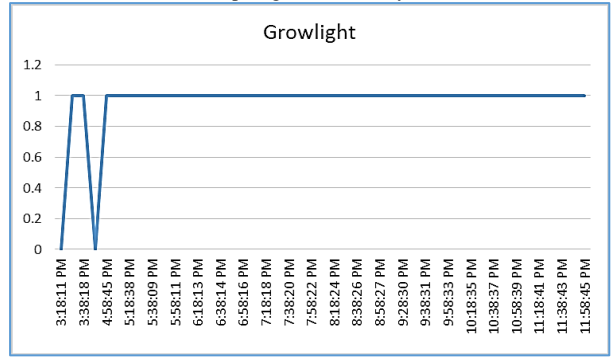

Figure 10: Graph of Sensor Data Log and Turn-On Time of Grow Light on Day 1

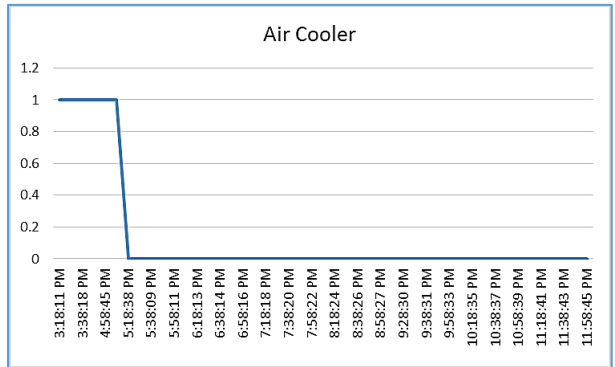

Figure 11: Graph of Sensor Data Log and Turn-On Time of Air Cooler on Day 1

The threshold values used by the proponents in regulating the system were 30°C for temperature, 300 ADC reading for soil moisture, and 5000 lux for light intensity. From the graphs it can be observed that whenever the values obtained by the sensors fall outside the range, the parameter regulators turns on, indicating that the system is working properly.

IV. CONCLUSION

On the technical aspect, the project is working properly because it was able to perform the functions required of it. The Arduino microcontroller was programmed correctly and was able to perform its required functions such as gathering data from the sensors and sending this data to the web database through the IoT gateway. The three sensors were able to measure precise values of the environmental parameters and the relay modules of the system were able to drive the parameter regulators based on the command it receives from the system, for automatic control, or the user, for the manual control. In addition, the Android application was able to provide the user with a means to remotely control and monitor the microfarm and the system was able to provide the user with a data log of the data from the sensors and the turn on and turn off times of the parameter regulators.

In the by-product aspect of the project, outstanding results has also been obtained since it was found out that the mustard green samples that were used in testing the effectivity of the system were about 0.23 cm taller than its standard height during the 27th day of growth.

This IoT-based microfarm system can be used to reduce human energy and effort involved in farming or gardening. Through this study, it has been proven to be an effective and economic way to reduce human effort, money and time wastage of time to time visitation. The researchers' goal of developing a microfarm system that can be monitored and controlled using a smartphone application has been met. This microfarm system can be used for urban/suburban farming and the developed system can fit in most small spaces such as garages or even indoors.

ACKNOWLEDGMENT

The authors would like to acknowledge Mr. Rolando A. Londonio of Pasay City Cooperative Development Office of the City Government of Pasay, Philippines under the leadership of its City Mayor, Hon. Antonino G. Calixto, for his very accommodating nature and for giving valuable insights and suggestions on how to improve the project, Mr. Rene Magdaraog of the Family-based Ecological Diversion and Recycling of Waste (FEDROW) for assisting the researchers throughout the deployment period of the project, and mentors and colleagues from the Electronics Engineering Department, College of Engineering, Technological University of the Philippines in Manila for the inspiration and backing in the entire course of leading this study.